\documentclass[useAMS,usenatbib]{mn2e}
\usepackage{color}
\usepackage{graphicx}
\usepackage{amssymb}
\usepackage{array}
\usepackage{color}
\usepackage{natbib}
\bibpunct{(}{)}{;}{a}{,}{,}

\newcommand{\degree}{\ensuremath{^\circ}}
\title[Magnetic field geometry of BRCs]{Magnetic field structure of IC\,63 and IC\,59 associated to H$~$II region - Sh 185}
\author[Soam et~al.]{A. Soam $^{1, 2}$\thanks{email:{archana@kasi.re.kr, archanasoam.bhu@gmail.com}}, 
Maheswar G.$^{2}$,
Chang Won Lee$^{1, 4}$,
Neha S.$^{2,3}$,
B-G Andersson$^5$
\\
$^{1}$ Korea Astronomy \& Space Science Institute (KASI), 776 Daedeokdae-ro, Yuseong-gu, Daejeon 305-348, Republic of Korea.\\
$^{2}$ Aryabhatta Research Institute of Observational Sciences (ARIES), Nainital, India - 263002\\
$^{3}$ Pt. Ravishankar Shukla University, Amanaka G.E.Road, Raipur, Chhatisgarh, India - 492010\\
$^{4}$ University of Science \& Technology, 176 Gajeong-dong, Yuseong-gu, Daejeon, Republic of Korea.\\
$^{5}$ SOFIA Science Centre, USRA, NASA Ames Research Centre, M.S. N211-3 Moffett Field, CA 94035, USA \\
}

\begin{document}
\date{Accepted------}

\pagerange{\pageref{firstpage}--\pageref{lastpage}} \pubyear{--}

\maketitle

\label{firstpage}

\begin{abstract}
Bright-rimmed clouds (BRCs) are formed at the periphery of H$~$II regions as the radiation from the central star interacts with dense gas. The ionization and resulting compression of the clouds may lead to cloud disruption causing secondary star formation depending on the stellar and gas parameters.  Here we use R-band polarimetry to probe the plane-of-the sky magnetic field in the two near-by BRCs IC\,59 and IC\,63. Both nebulae are illuminated by $\gamma$ Cas with the direction of ionizing radiation being orientated parallel or perpendicular to the local magnetic field, allowing us to probe the importance of magnetic field pressure in the evolution of BRCs.  Because of the proximity of the system ($\sim$200pc) we have acquired a substantial sample of over 500 polarization measurements for stars background to the nebulae. On large scales, the magnetic field geometries of both clouds are anchored to the ambient magnetic field. For IC 63, the magnetic field is aligned parallel to the head-tail morphology of the main condensation, with convex morphology relative to the direction of the ionizing radiation. We estimate the plane of the sky magnetic field strength in IC\,63 to be $\sim90\mu$G. In IC\,59, the projected magnetic field follows the M shape morphology of the cloud.  Here, field lines present a concave shape with respect to the direction of the ionizing radiation from $\gamma$ Cas. Comparing our observations to published theoretical models we find good general agreement, supporting the importance of magnetic fields in BRC evolution.
\end{abstract}

%#############################             INTRODUCTION

\section{Introduction}
When a massive star is born in a clumpy molecular cloud, the clumps that become exposed to the ionizing radiation coming from the massive star will begin to photoevaporate. The ionization heating generates a shock which will converge into the cloud compressing it to form single or multiple cores. These core(s) may eventually collapse to form stars. Several patches of obscuration are often noticed towards the periphery of relatively matured H$~$II regions \citep[e.g., ][]{1983ApL....23..119Z, 1991ApJS...77...59S, 1994ApJS...92..163S, 1996AJ....111.2349H, 1998PASA...15...91O, 2010A&A...523A...6D, 2010A&A...518L..81Z, 2012AJ....144..173D, 2012A&A...542L..18S}. The patches that show bright rim on the surface facing the central ionizing source(s) produced by the recombination radiation from ionization front (I-front) are termed as bright-rimmed clouds (BRCs). \citet{1991ApJS...77...59S} and \citet{1994ApJS...92..163S} have cataloged 89 BRCs that are found to be associated with the  Infrared Astronomical Satellite (IRAS) point sources (44 in the northern and 45 in the southern hemisphere). BRCs were initially categorized into types A, B and C according to their length-to-width ratios with type C being the most elongated ones \citep{1991ApJS...77...59S, 1994ApJS...92..163S}. A large number of these BRCs show signs of ongoing  star formation \citep{1989ApJ...342L..87S, 1995ApJ...455L..39S, 2002AJ....123.2597O, 2011MNRAS.415.1202C, 2012PASJ...64...96H, 2012Ap.....55..471M} which may have possibly been triggered due to the ionization front formation in these clouds. Thus BRCs are the ideal laboratories to investigate the radiation driven implosion (RDI) process which is suggested as the mechanism responsible for the compression of the cloud and subsequent triggering of star formation \citep{1983A&A...117..183R, 1989ApJ...346..735B, 1994A&A...289..559L}. 

Several analytical \citep[e.g., ][]{1955ApJ...121....6O, 1968Ap&SS...1..388D, 1969Phy....41..172K}, semi-analytical and numerical modelling studies have been carried out over the past few decades to follow the details of the RDI of photionized clump and its subsequent acceleration away from the ionizing source \citep[e.g., ][]{1981A&A....99..305T, 1982ApJ...260..183S, 1983ApJ...271L..69K, 1984A&A...135...81B, 1989ApJ...346..735B, 1990ApJ...354..529B, 1994A&A...289..559L, 2003MNRAS.338..545K, 2009ApJ...694L..26G, 2009ApJ...692..382M, 2011ApJ...736..142B, 2012MNRAS.420..562H, 2013MNRAS.431.3470H, 2015MNRAS.450.1017K}. Significant differences in the values of the allowed velocities of the propagation of the ionization fronts were seen between magnetized and non-magnetized cases when the continuity equations were solved by taking a simple case of magnetic field parallel \citep{1998A&A...331.1099R} and oblique \citep{2000MNRAS.314..315W} to the ionization front. The first 2-D radiation-magnetohydrodynamics (R-MHD) calculation was performed by \citet{2007Ap&SS.307..179W}. Fields of various strengths and orientations were considered to show the effects of magnetic field in the structural evolution of clumps interacting with ionizing radiation. By including magnetic field of various strengths and orientations, \citet{2009MNRAS.398..157H} carried out 3-D R-MHD simulations. The results show that while a strong perpendicular (with respect to the direction of ionizing radiation) magnetic field orientation creates a flattened, plate-like structure, an oblique initial field results in to a comma-shaped cloud morphology.  Thus it is important to trace the cloud scale magnetic field of BRCs observationally to verify the results obtained in the above models.

Observations of starlight polarization by absorption in visible and near-infrared wavelengths are used as a tool to trace the plane of the sky component of the magnetic field \citep{1949Sci...109..166H, 1949ApJ...109..471H, 1949Sci...109..165H}. This technique has been used extensively in several studies to derive the magnetic field maps towards the molecular clouds in different environments \citep[e.g., ][]{1976AJ.....81..958V, 1987ApJ...319..842H, 1992MNRAS.257...57B, 2002ApJS..141..469P, 2011ApJ...741...21C, 2011ApJ...734...63S, 2013MNRAS.432.1502S, 2014A&A...565A..94B, 2015ApJ...798...60K, 2015ApJ...803L..20S}. Polarization of starlight at visible and near-infrared wavelengths is due to the diachroism by non-spherical dust grains. These grains are aligned with their short axis parallel to the local magnetic field \citep{2003dge..conf.....W}. The dichroism of the extinction along the long axis results in a net polarization which is parallel to the magnetic field. The same aligned dust grains can emit polarized thermal radiation in the sub-millimeter and millimeter wavelengths \citep[e.g.,][]{1982MNRAS.200.1169C, 1998ApJ...502L..75R, 2000ApJS..128..335D, 2010ApJS..186..406D, 2012ApJS..201...13V, 2014ApJS..213...13H}. The net polarization of radiation emitted by dust grains is parallel to the long axis of the grains hence has to be rotated by 90$\degree$ to get the sense of projected field orientation. The actual mechanism by which the dust grains align with the magnetic field in a manner described above is still unclear \citep{2007JQSRT.106..225L, 2015ARA&A..53..501A}. However, radiative torque mechanism, originally proposed by \citet{1976Ap&SS..43..291D}, seems to be emerging as the most successful one in explaining the dust alignment in various environments \citep[e.g., ][]{2014MNRAS.438..680H, 2015MNRAS.448.1178H, 2015ARA&A..53..501A}.

Magnetic field maps have so far been acquired for only four out of eighty nine BRCs, namely: BRC 20 \citep{2011ApJ...743...54T}, BRC 51 \citep{2011ApJ...743...54T, 1999MNRAS.308...40B}, BRC 74 \citep{2015ApJ...798...60K} and BRC 89 \citep{2014ApJ...783....1S}. Based on a small number of stars, \citet{1999MNRAS.308...40B} suggested that the magnetic field in CG 30-31 complex (which includes BRC 51) is parallel to the tail. Cometary Globules (CGs) are molecular clouds that show a compact bright-rimmed head and a faint tail geometry that extends from the head and points away from a nearby photoionizing source. However, the magnetic fields mapped by \citet{2011ApJ...743...54T} in CG 30-31 show a rather complex geometry. The aim of their work was to investigate a possible correlation between the interstellar magnetic field and the protostellar jets. Combining the results from both the studies, it seems like the field lines are following the curvature of the head and becoming parallel towards the tail of the CG 30. The polarization measurements towards BRC 20 (OriI-2) are not sufficient to draw any conclusion. The magnetic field geometry in BRC 74 and BRC 89 are found to be symmetric about the cloud axis. The magnetic field geometry as a whole shows two distinct almost orthogonal field orientations. While one of the components is running along the rim of the cloud perpendicular to the direction of the ionizing photons, the other component is running parallel to the direction of the ionizing radiation (similar to the trend seen in CG 30-31 complex). Using near infrared polarization of background stars, \citet{2007PASJ...59..507S} traced the magnetic field geometry of the Eagle nebula (M16) region. The field lines inside some of the pillar structures in M16 were found to be aligned in the direction of the exciting star. But they were quite different from the ambient magnetic field geometry surrounding them. This clearly indicates that the magnetic fields inside the pillars have been affected by the dynamical impact of the ionizing radiation. \citet{2011MNRAS.412.2079M}, explaining the results seen towards M16, showed by simulations that an initially weak and medium strength magnetic field oriented perpendicular to the direction of ionizing radiation can be made to align with it during the dynamical evolution of the cloud.
 
In order to study the effects of magnetic fields on the structural evolution of clouds, it is appropriate to select H$~$II regions (a) that contain multiple BRCs or clouds located at different positions around the ionizing source, (b) where the properties of the ionizing source(s) are known, (c) that are located in close proximity with less foreground extinction, and (d) that show a very simple geometry. If we assume that the magnetic field had a preferred orientation prior to the formation of an H$~$II region, then clouds located at different positions around it are expected to evolve differently because of the difference in the relative orientation between the direction of the ionizing radiation and the initial magnetic field orientation. 

In this study, we made $R$-band polarization measurements of over 500 stars projected on the Sharpless 185 H$~$II region \citep[Sh 185; ][]{1959ApJS....4..257S}. This region is associated with two nebulae, IC\,59 and IC\,63. IC\,63 was classified as a type-B BRC by \citet{1991ApJS...77...59S} and \citet{2010ApJ...717..658M} proposed a M-type classification for IC\,59. Sh 185 is illuminated and ionized by a B0 IV star, $\gamma$ Cas \citep{2005AJ....129..954K}. Based on the new parallax measurements by \citet{2007A&A...474..653V},  the estimated distance to this star is $168\pm4$ pc making it one of the closest H$~$II regions to the Sun. 

In this paper, section 2 represents the observations and techniques used for the data reduction and analysis. Our results are presented in section 3. We discuss our results in section 4 and conclude with a summary in section 5.

%==============
%\input{table2}		%Observation Log
%==============
%##############################                OBSERVATIONS AND DATA REDUCTION
\section{POLARIZATION OBSERVATIONS AND REDUCTION}\label{sec:obs,datared}
The optical polarimetric data towards IC 63 and IC 59 were collected using the Aries IMaging POLarimeter (AIMPOL) \citep{2004BASI...32..159R} mounted at the Cassegrain focus of the 104-cm Sampurnanand telescope, India. The images were obtained using a 1024$\times$1024 pixel$^{2}$ CCD chip (Tektronix TK1024) out of which the central 325$\times$325 pixel$^{2}$ area was used for the imaging. The plate scale of the CCD is 1.48 arcsec$/$pixel and the field of view (FOV) is $\sim 8$ arcmin. The mean exposure time per field per half-wave plate (HWP) position was $\sim$250 s. The observed stellar image varied from 2 to 3 pixels. The read out noise and gain of CCD were 7.0 $e^{-1}$  and 11.98 $e^{-1}$/ADU, respectively. A R-band filter, matching the Kron-Cousin pass-band ($\lambda_{eff}$=0.760$\mu$m) band was used to observe 11 fields covering IC 63 and IC 59 nebulae. The targets to be observed were selected using Wide-field infrared survey explorer \citep[WISE; ][]{2010AJ....140.1868W}12$\mu$m images.

%The observing lines-of-sight were visually selected from inspection of the WISE 12$\mu$m emission image of the these nebulae from Sky View. The outskirts of relatively high and low density regions were selected. 

AIMPOL works in linear polarization mode and consists of an achromatic HWP modulator and a Wollaston prism beam-splitter. This arrangement provides two images of each object on the CCD. A half-wave retarder is rotated to obtain four normalized Stokes parameters, q[R($0\degree$)], u[R($22.5\degree$)], $q_{1}$[R($45\degree$)] and $u_{1}$[R($67.5\degree$)] corresponding to its four positions i.e. $0\degree$, $22.5\degree$, $45\degree$ and $67.5\degree$. We estimate the errors in normalised Stokes parameters ($\sigma_R$)($\alpha$)($\sigma_q$, $\sigma_u$, $\sigma_{q1}$, $\sigma_{u1}$) in per cent using the relation given by \citet{1998A&AS..128..369R}.  We have also estimated the average background counts around the extraordinary and ordinary rays \citep{1998A&AS..128..369R}. 

After the normal steps of CCD reductions (i.e. flat-fielding using flux normalization formula from \citet{1998A&AS..128..369R}, bias subtraction, aligning and combining the frames), we identified the corresponding pairs of stars and performed photometry on them in each of the field using the IRAF DAOPHOT package. From the obtained file containing magnitude data, we calculate the polarization values. A set of programs written in IDL and python languages for this purpose reads the files containing fluxes of the stars and estimate the degree of linear polarization (P in per cent) and polarization angle ($\theta_{P}$, measured from the north to the east in degree).
 
 In order to check the instrumental polarization, we observed zero-polarization standard stars during every run. The typical instrumental polarization is found to be less than $\sim0.1\%$ in AIMPOL \citep[see ][]{2013MNRAS.432.1502S, 2014Ap&SS.350..251S}. The reference direction of the polariser was determined by observing polarized standard stars from \citet{1982ApJ...262..732H} and \citet{1992AJ....104.1563S}. The measured calibration results are shown in Table \ref{tab:std}. For all observing seasons, the observed position angle values were in good agreement with the standard values given in the literature (see Table 2). The zero point offset was corrected on every run using the offset seen between the standard position angle values given in \citet{1982ApJ...262..732H} and \citet{1992AJ....104.1563S} and those obtained by us.

%*************************************************************************
%*************************************  TABLE 1 ************************** **************************************************************************
\begin{table}
\caption{Log of polarization observations in $R_{kc}$ filter.}\label{tab:obslog}
\begin{tabular}{p{1.3cm}p{6cm}}\hline
 Cloud ID           &  Date of observations (year, month,date)\\
\hline
 IC\,59/63          & 2013, Nov, 06,08,09,28 \\ 
                    & 2013, Dec, 01, 02, 03 ,28 ; 2014, Jan 01 \\
\hline
\end{tabular}
\end{table}
%*************************************************************************  

%************************table std************************************************
%\input{table3}
\begin{table}
\caption{Results of observed polazised standard stars.}\label{tab:std}
\begin{tabular}{lll}\hline
Date of     &P $\pm$ $\sigma_P$ 	&  $\theta$ $\pm$ $\sigma_{\theta}$  \\
Obs.		&(\%)            		& ($\degree$)                           \\\hline
\multicolumn{3}{l}{{\bf HD\,236633}}\\
\multicolumn{3}{l}{$^\dagger$Standard values: 5.38 $\pm$ 0.02\%, 93.04 $\pm$ 0.15$\degree$}\\
01 Dec 2013 & 5.5 $\pm$ 0.1     & 92 $\pm$ 1\\
01 Jan 2014	& 4.8 $\pm$ 0.2     & 92 $\pm$ 1 \\\hline
\multicolumn{3}{l}{{\bf HD\,25443 }}\\
\multicolumn{3}{l}{$^\dagger$Standard values: 4.73 $\pm$ 0.05\%, 133.65$\pm$ 0.28$\degree$}\\
08 Nov 2013 & 4.9 $\pm$ 0.1     & 133 $\pm$ 2 \\
09 Nov 2013 & 4.9 $\pm$ 0.1     & 132 $\pm$ 1  \\ 
02 Dec 2013 & 4.7 $\pm$ 0.1     & 133 $\pm$ 1  \\ 
03 Dec 2013 & 4.8 $\pm$ 0.1     & 132 $\pm$ 2  \\ 
28 Dec 2013 & 4.7 $\pm$ 0.1     & 132 $\pm$ 1  \\ \hline
\multicolumn{3}{l}{{\bf BD$+$64$\degree$106}}\\
\multicolumn{3}{l}{$^\dagger$Standard values: 5.69 $\pm$ 0.04\%, 96.63 $\pm$ 0.18$\degree$}\\
01 Dec 2013 & 5.3 $\pm$ 0.1     & 96 $\pm$ 1   \\\hline
\multicolumn{3}{l}{{\bf HD\,19820}}\\
\multicolumn{3}{l}{$^\dagger$Standard values: 4.53 $\pm$ 0.02\%, 114.46$\pm$ 0.16$\degree$}\\
02 Dec 2013 & 4.5 $\pm$ 0.1     & 113 $\pm$ 1  \\
03 Dec 2013 & 4.5 $\pm$ 0.1     & 113 $\pm$ 1  \\\hline
\multicolumn{3}{l}{{\bf HD\,43384}}\\
\multicolumn{3}{l}{$^\ddagger$Standard values: 4.89 $\pm$ 0.03\%, 59.10 $\pm$ 0.17$\degree$}\\\hline
28 Nov 2013 & 4.7 $\pm$ 0.1     & 57 $\pm$ 3  \\ \hline
\end{tabular}

$^\dagger$ In $R_{c}$ band from \citet{1992AJ....104.1563S} \\
$\ddagger$ Values obtained from \citep{1982ApJ...262..732H} \\
\end{table}
%=====================================36===============================================

\section{Results}\label{result}

%-----------------------------------------------------------------
\begin{figure*}
%\resizebox{18cm}{9.5cm}{\includegraphics{combined_fracp2.5_23June16_W12.eps}}
%\resizebox{18cm}{9.5cm}{\includegraphics{fig1_combined_ic6359_W12.eps}}
%\resizebox{18cm}{9.5cm}{\includegraphics{\mypath/fig1_combined_ic6359_W22.eps}}
\resizebox{17cm}{9.0cm}{\includegraphics{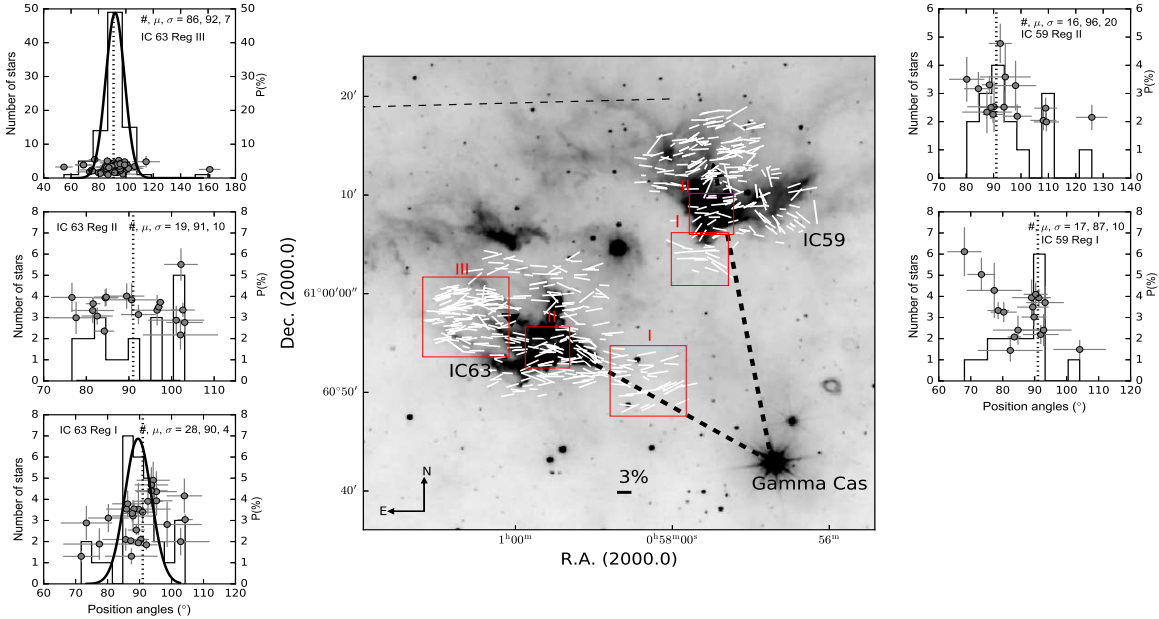}}
\caption{The orientation of the polarization position angles measured for the stars projected towards IC\,59 and IC\,63 clouds are drawn on the 48$^{\prime}\times48^{\prime}$ WISE 12$\mu$m image. The nebulae, IC\,59 and IC\,63 and the ionizing star, $\gamma$ Cas, are identified and labelled. The length of the vectors correspond to the measured P. A 3\% vector is drawn for comparison at the lower left corner. The arrows with thick broken-line show the direction of the ionizing radiation. The arrows pass through the densest part of the nebulae identified from the 12$\mu$m intensity. The broken-line shown on the top left corner represents the inclination of the Galactic plane. The North and East directions are shown in the figure. The gaussian fitted histograms with the distribution of P and $\theta_{P}$  corresponding to the regions marked in IC\,63 and IC\,59 are also shown.}\label{Fig:combined_IC6359}
\end{figure*}
%-----------------------------------------------------------------

We obtained polarization measurements of 543 stars towards the Sh 185 region. Of those, 307 and 236 stars are from IC\,63 and IC\,59 regions, respectively. We selected only those sources in our analysis for which signal to noise ratio P/$\sigma_P\geqslant2.5$. Our polarization results are shown in Fig. \ref{Fig:combined_IC6359}. In this figure, a 48$^{\prime}\times48^{\prime}$ WISE 12$\mu$m image shows the two nebulae, IC\,63 and IC\,59 and the ionizing star, $\gamma$ Cas. The orientation and the length of the vectors correspond to the measured $\theta_P$ and P (in per cent), respectively ($\theta_P$ values are measured from the north increasing eastward). The arrowed thick broken-lines show the projected direction of the ionizing radiation. The arrows are drawn from the position of $\gamma$ Cas and made to pass through the position where the 12$\mu$m intensity is found to peak in the nebulae. The 12$\mu$m intensity peak coincides with the one identified from CO observations in IC\,63 \citep{1994A&A...282..605J}. 

\begin{figure}
%\resizebox{8cm}{15.0cm}{\includegraphics{\mypath/theta_Histogram.eps}}
%\resizebox{8cm}{15.0cm}{\includegraphics{IC63_histgauss_and_P_PA.eps}}
\resizebox{8cm}{15.0cm}{\includegraphics{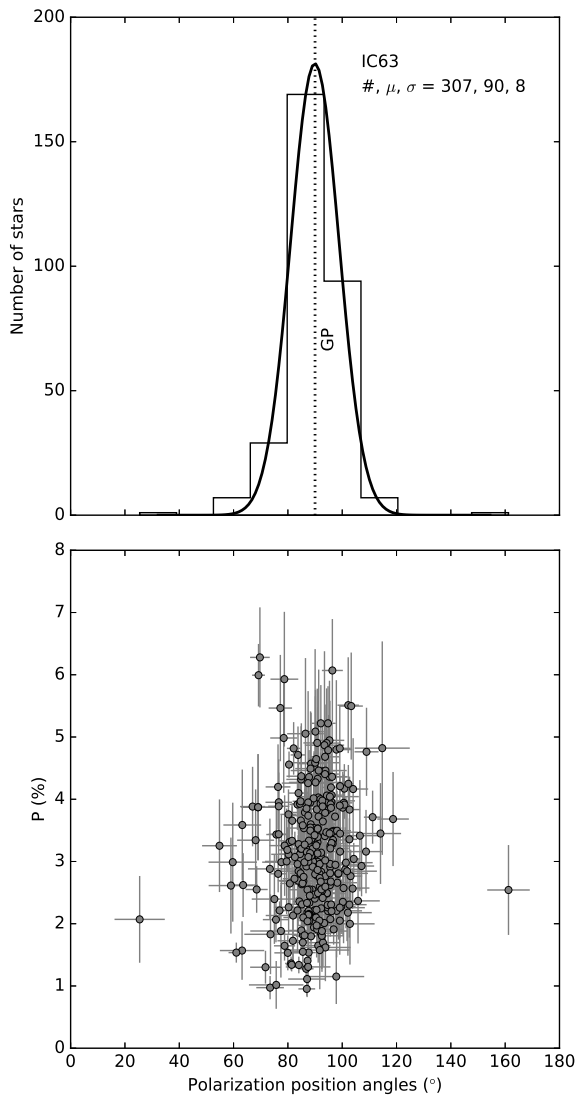}}
\caption{{\bf Upper panel}: Gaussian fitted histogram of $\theta_{P}$ with bin size $10\degree$ in IC\,63. The orientation of Galactic plane at the latitude of the cloud is shown using dotted line. {\bf Lower panel}: Variation of P with $\theta_{P}$ of program stars observed towards IC\,63.}\label{Fig:hist_IC63}
\end{figure}

\begin{figure}
%\resizebox{8cm}{15.0cm}{\includegraphics{\mypath/theta_Histogram.eps}}
%\resizebox{8cm}{15.0cm}{\includegraphics{IC59_histgauss_and_P_PA.eps}}
\resizebox{8cm}{15.0cm}{\includegraphics{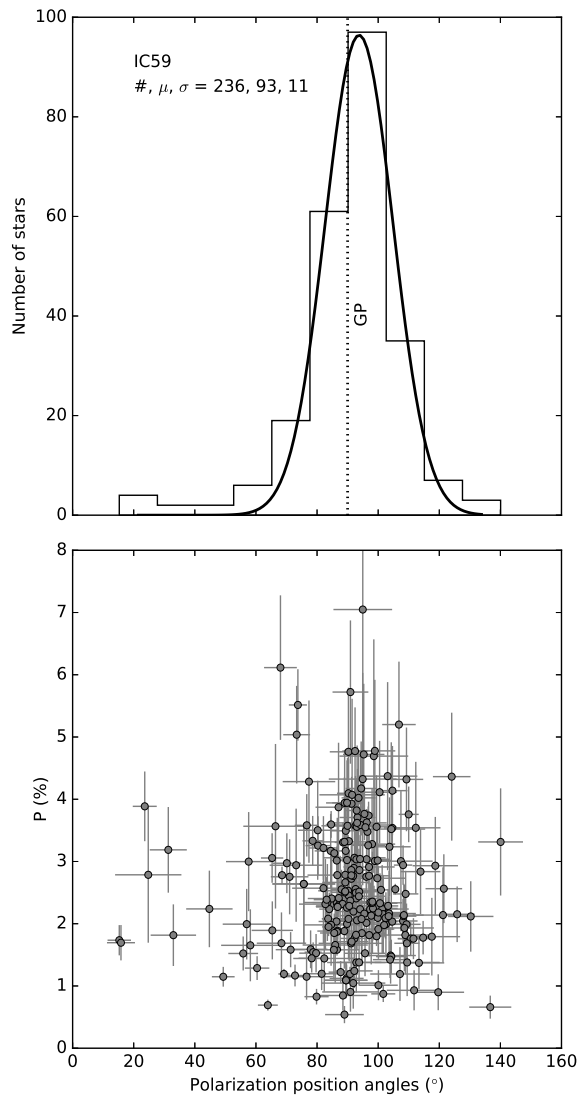}}
\caption{{\bf Upper panel}: Gaussian fitted histogram of $\theta_{P}$ with bin size $10\degree$ in IC\,59. The orientation of Galactic plane at the latitude of the cloud is shown using dotted line. {\bf Lower panel}: Distribution of P with $\theta_{P}$ of program stars observed towards IC\,59.}\label{Fig:hist_IC59}
\end{figure}

The histogram of polarization position angles towards IC\,63 and IC\,59 are shown in Fig. \ref{Fig:hist_IC63} and Fig. \ref{Fig:hist_IC59}, respectively. The lower panels of these figures show the distribution of P versus $\theta_{P}$. The P values in IC\,63 are found to range from 0.9\% to 7.0\% and those in IC\,59 range from 0.5\% to 6.2\%. The mean value of P with corresponding standard deviation is found to be 3.1$\pm$1.1\% and 2.5$\pm$1.1\% in IC\,63 and IC\,59, respectively. The mean and the standard deviation values of $\theta_{P}$ obtained from a Gaussian fit to the distribution are found to be 90$\pm$8$^\circ$ and 93$\pm$11$^\circ$ in IC\,63 and IC\,59, respectively. Gaussian fitted histograms of $\theta_{P}$ in the different regions of IC\,63 and IC\,59 are shown in Fig. \ref{Fig:combined_IC6359}.

%-----------------------------------------------------------------
%-----------------------------------------------------------------

\section{Discussion}\label{sec:disc}

\subsection{The distance and the ambient magnetic field}\label{sec:distance}

%-----------------------------------------------------------------
\begin{figure}
%\resizebox{9.4cm}{14.0cm}{\includegraphics{P_PA_vs_dist_heiles_py.eps}}
\resizebox{9.4cm}{14.0cm}{\includegraphics{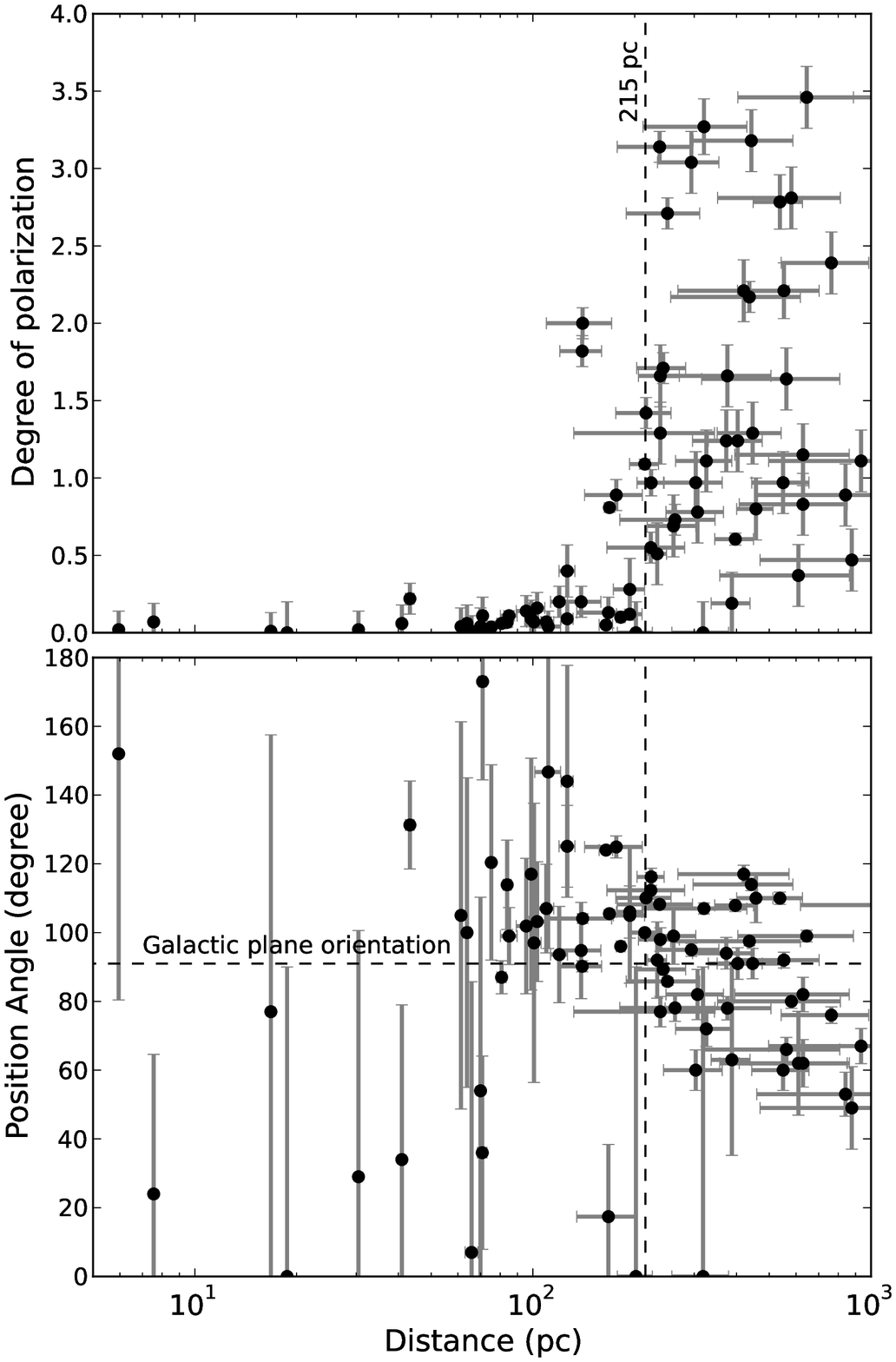}}
\caption{Polarization results for stars from a region of 10$\degree$ radius around $\gamma$ Cas obtained from the \citet{2000AJ....119..923H} catalog. The distance of the stars are calculated using the parallax measurements made by the Hipparcos \citep{2007A&A...474..653V}. In the upper panel, we plot the P$\%$ values as a function of distance. The vertical broken line is drawn at $\sim215\pm40$ pc where a sudden rise in the P$\%$ values are noticed. In the lower panel, we show the polarization position angles (in degree) as a function of distance for the same stars. The horizontal broken line drawn at $\sim90\degree$ shows the orientation of the Galactic plane at the location of $\gamma$ Cas.}\label{Fig:dist_pol_sh185}
\end{figure}
%-----------------------------------------------------------------

Based on the spectrophotometric results obtained from the literature, \citet{1982ApJS...49..183B} estimated $210\pm70$ pc as the distance to Sh 185. Accounting for its variability in magnitude and color, \citet{1984PASJ...36..231V} evaluated a distance of $230\pm70$ pc to the ionizing star $\gamma$ Cas. By taking the Hipparcos parallax measurements of $\gamma$ Cas from \citet{1997A&A...323L..49P}, the star was placed at a distance of $188\pm20$ pc in earlier studies. However, if we take re-calculated Hipparcos parallax value of $\gamma$ Cas from \citet{2007A&A...474..653V}, the distance to the star becomes $168\pm4$ pc. \citet{2013ApJ...775...84A}, by estimating extinction of field stars with trigonometric or spectroscopic parallaxes, estimated a distance of $\sim$200 pc to Sh 185 region. 

The polarization and the parallax measurements available for stars towards Sh 185 are used here to estimate the distance. As mentioned earlier, since polarization is produced due to the differential extinction of the light by the dust grains lying along the pencil-beam of the star, similar to the extinction, as the column of the dust increases along the pencil-beam, the degree of polarization also shows a systematic rise. This is true provided there is no change in the magnetic field orientation along the path of the light. When the light passes through a dust cloud or a dust layer at some distance, the polarization degree values measured for the background stars are expected to show an abrupt rise. The distance at which the sharp rise occurs can be taken as the distance to the cloud or the dust layer, if the distance distribution of the stars is assumed to be uniform. This technique was used in the past to estimate distances to interstellar clouds \citep[e.g.,][]{2007A&A...470..597A}. From a circular region of 10$\degree$ around $\gamma$ Cas, we selected stars for which both polarization \citep[from the ][ catalog]{2000AJ....119..923H} and the Hipparcos parallax measurements \citep{2007A&A...474..653V} are available. The 10$\degree$ circular region is taken to obtain a good sample of stars to constrain the distance better. The degree of polarization versus distance plot is shown in Fig. \ref{Fig:dist_pol_sh185} (upper panel). A sharp rise in the values of P is found to be occurring at $\sim215$ pc. The average value of P for stars located at distances $<215$ pc is $\sim0.3\%$ whereas that for stars located at $\geqslant215$ pc is $\sim1.6\%$. The uncertainty in the distance quoted here is that of a star which is located at this distance. The two stars showing $\sim2\%$ polarization at distances less than $\sim215$ pc are identified as HD 12340 (B9 III) and HD 12708 (B9 V). The distance found by the hipparcos parallax values keeps these two stars in front of the cloud. But the photometric distance of HD12340 and HD12708 calculated using their extinction values are found to be $\sim$700 pc and $\sim$320 pc, respectively. Surprisingly, there is a huge difference between the distance values estimated from parallax and the photometric method. The larger photometric distances may be the cause attributed to their relatively higher degree of polarization. Low values of P seen for the stars located at distances $\lesssim215$ pc suggest that there are negligible amount of foreground obscuration towards Sh 185 region. 

The polarization position angles are plotted against the distances of the stars in Fig. \ref{Fig:dist_pol_sh185} (lower panel). The orientation of the Galactic plane at the location of $\gamma$ Cas is shown by a broken horizontal line ($\sim90\degree$). The weighted average of the $\theta_{P}$ for the stars located $<215$ pc is found to be $110\degree$. We take this as the orientation of the initial magnetic field prior to the formation of Sh 185.  A systematic change in the values of $\theta_P$ from $\sim100\degree$ to $\sim60\degree$ is apparent for distances larger than $\geqslant 215$ pc. The weighted average of the $\theta_P$ of the stars located $\geqslant 215$ pc is found to be $95\degree$. It is clear from the data presented here that the large scale magnetic field which represents region around Sh 185 is oriented roughly parallel to the Galactic plane. This is consistent to the results of the $\lambda~6~$cm polarization survey of Galactic plane by \citet{2007A&A...463..993S} who noted a very uniform large-scale magnetic field running parallel to the Galactic plane. The cloud magnetic field parallel to the Galactic plane implies that the large scale Galactic magnetic fields anchors the clouds.

%that in the large-scale structures, the polarization angles are concentrated around 0$^{\degree}$ indicating 
%

As discussed above, the foreground obscuration towards Sh 185 is very low (the average value of P for the stars located at distances $<215$ pc is $\sim0.3\%$). Even for an optimum alignment of the dust grains, using the relation, P/A$_{V}$ $\approx$ 3\%, the maximum foreground extinction would be only $\sim$0.1 magnitude. For this reason, we have not removed any contribution due to the foreground material from our observed polarization values. The mean of the $\theta_P$ values  of the two nebulae are found to be consistent with that obtained for the stars taken from the \citet{2000AJ....119..923H} catalog from a wider area towards Sh 185. This implies that the magnetic fields of the two nebulae are anchored to the field from the surrounding interstellar medium. However, there are magnetic fields structures seen at the smaller scales of the clouds.

\subsection{IC\, 63}

\begin{figure}
%\resizebox{9.5cm}{8.0cm}{\includegraphics{IC63_only_fracp0.4_zoomed_mod_W12.eps}}
%\resizebox{9.5cm}{8.0cm}{\includegraphics{\mypath/IC63_only_fracp0.4_zoom_3Jun16_W22.eps}}
\resizebox{9.5cm}{8.0cm}{\includegraphics{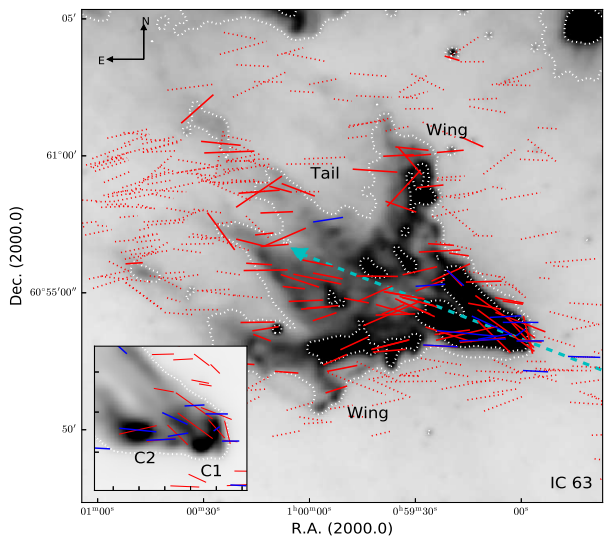}}
\caption{The polarization results of 307 stars observed towards IC\, 63 region are shown on the 20$^{\prime}\times 20^{\prime}$ WISE 12$\mu$m image. The vectors drawn inside the outer most contour drawn are shown using thick red lines. The ones outside this contour is shown using red dotted-lines. The broken arrow in cyan shows the direction of the ionizing photons from $\gamma~$Cas. The vectors drawn in blue are the results are obtained from \citet{2013ApJ...775...84A}. The North and East directions are shown in the figure.}\label{Fig:IC63vectors}
\end{figure}

The polarization vectors showing the plane of the sky component of the magnetic field geometry of IC\,63 are presented in Fig. \ref{Fig:IC63vectors}. The vectors show a well ordered magnetic field geometry over the nebula which is consistent with a very low dispersion seen in the polarization angles. The vectors  shown in thick red lines are those that are located within the outer most contour drawn at 0.29 and 2.03 Jy\,sr$^{-1}$ of the WISE 12$\mu$m intensity (Fig.  \ref{Fig:IC63vectors}). The nebula shows an extended head structure which is more sharply defined with its apex pointing directly towards the  ionizing star, $\gamma$ Cas. There is a bright rim towards the side facing this star. Behind and to the northern and southern sides of the protruding head, there are two wing-like structures (as identified in Fig. \ref{Fig:IC63vectors}). The arrowed broken line in cyan shows the projected direction of the ionizing radiation from the $\gamma$ Cas \citep[$\sim60\degree$ with respect to the north;][]{2013ApJ...775...84A}. 

The vectors lying outside the cloud boundary are oriented almost along $\sim90\degree$ (shown using red dotted vectors). The magnetic field geometries in three different regions (reg I: outside the cloud near the head part, reg II: on the head part of the cloud and reg III: outside the cloud near tail part) shown by red boxes towards IC\,63 are indicated in Fig. \ref{Fig:combined_IC6359}. The gaussian fitted histogram of $\theta_{P}$ alongwith its distribution with degree of polarization in these regions are also shown. The distribution of $\theta_{P}$ in the region lying on the head part seems to be more dispersed. The magnetic field geometry of the head part alone is shown separately in the inset (Fig. \ref{Fig:IC63vectors}). Two distinct condensations (C\,1 and C\,2) are clearly visible towards the head part of IC\,63. The field lines around C1 appear to trace the elongated material lying behind the bright rim and oriented almost along the direction of the ionizing radiation. These condensations could be the fragments due to the uneven evaporation rate and acceleration experienced by the density inhomogeneities existed in the cloud prior to its exposure to the ionizing radiation. \citet{2014MNRAS.444.1221K} investigated the evolution of prolate clouds on the boundaries of H$~$II regions using Smoothed Particle Hydrodynamics (SPH) simulations. They suggested that various fragmented cores found on the boundaries of H$~$II regions may result from the interaction between the ionizing radiation and preexisting prolate clouds of different initial geometrical and physical conditions. Also in the case of  IC\,63 the two condensations could be the result of fragmentation of the shocked surface layer of the cloud, due to gravitational instability. On a slightly larger scale, the field geometry is convex with respect to the projected direction of the ionizing radiation. As we move further towards the east of the head, the field lines continue to thread along the boundary of the wing structure, especially the structure to the southern side of  IC\,63. Towards the tail part, the fields lines are becoming more parallel to the direction of the ionizing radiation.

In near-infrared, the polarization of background starlight can trace the magnetic field geometry of regions with relatively higher extinction \citep{2011ASPC..449..207T, 2015ApJS..220...17K}. \citet{2013ApJ...775...84A} made measurements of visual extinction and H-band polarization values in a number of stars towards IC\, 63. Figure 3 of \citet{2013ApJ...775...84A} shows the measured H-band polarization of background starlight projected on and outside the cloud periphery shown on WISE image of IC\,63. The mean value of the polarization position angles obtained by \citet{2013ApJ...775...84A} is $90\degree$.

The projected angular offset between the magnetic field orientation in  IC\,63 and the direction of the ionizing radiation from $\gamma$ Cas is found to be $\sim30\degree$. This is by assuming that both $\gamma$ Cas and IC\, 63 nebula are located at same distance. However, evidence from earlier studies suggests that this may not be the case. For instance, the width of the ionized gas rim seen in  IC\,63 is  0.015 pc \citep{2005AJ....129..954K} which is a factor of three higher than the expected width (0.002 pc) of the ionized region inside the cloud considering the cloud density and the incident photon flux from $\gamma$ Cas. This value (0.002 pc) is predicted theoretically by \citet{2005AJ....129..954K} using CLOUDY model \citep{2002hbic.book.....F}. \citet{2005AJ....129..954K} suggested that an angle of inclination of 10$\degree$ with respect to the line of sight could produce the observed width in the nebula. In the chemical analysis of  IC\,63 using various models, \citet{1995A&A...302..223J, 1996A&A...309..899J} hypothesized that a large inclination of IC\,63 would be required to explain the anomaly of the extended C$^{+}$ and C emission observed towards the tail of the cloud. \citet{1996A&A...309..899J} concluded that the strong C$^{+}$ emission towards the tail part of IC\,63 indicates that the star $\gamma$ Cas is somewhat in front of the cloud, illuminating it not only edge-on, but also partially face-on. \citet{2013ApJ...775...84A}, based on the orientations of the H$_{2}$ streamers that they identified and the position angle of the Rayleigh scattering component seen towards one of the stars studied by  them, estimated an inclination of 58$\degree$ for IC\, 63 with respect to the plane of the sky. By considering the kinematics of  various components of gas towards the region with respect to that of $\gamma$ Cas, they suggested that the location of the nebula to the far side of the $\gamma$ Cas is the most plausible scenario. 
%Using a dynamical approach, in the next section, we have made an attempt to examine the inclination of the cloud with respect to the plane of the sky.

%-----------------------------------------------------------------

%\subsubsection{Cloud parameters and its inclination with respect to the plane of the sky}
\subsubsection{Cloud parameters}

When a neutral cloud is exposed to the ionizing radiation from an O- or B-type star, the incident flux of energetic photons ionizes the surface of the cloud. The ionized gas streams radially away from the I-front. The pressure exerted by the ionized gas at the surface of the cloud drives a shock front which compresses the cloud causing it to implode.  Using two dimensionless parameters, $\eta$ and $\nu$, \citet{1989ApJ...346..735B} classified clouds that are subjected to ionizing radiation. While $\eta$ represents the initial column density of the cloud, $\nu$ is a measure of the initial I-front driven shock velocity. Making use of the expressions given by \citet{1989ApJ...346..735B}, we calculated the parameters $\eta$ and $\nu$ for IC\,63. We took 7.9$\times10^{47}$ photons per second as the rate of photons beyond the Lyman limit for the $\gamma$ Cas. We obtained this value from \citet{1996ApJ...460..914V} for a star of B0.5V spectral type. As pointed out by \citet{2004A&A...426..535M}, there exists a range of spectral type designations for $\gamma$ Cas. But in order to be consistent with most of the previous studies, we also have adopted the spectral type that is given in the SIMBAD database.  Reduction in the photon flux due to the absorption by the intervening material between $\gamma$ Cas and IC\,63 is considered to be negligible. The radius of the cloud is taken as 0.03 pc by fitting a circle to the head part of IC\,63 and the number density is taken as $(5\pm2)\times10^{4}$\,cm$^{-3}$ from \citet{1994A&A...282..605J}. Our estimated radius is found to be consistent with the extent of gas and dust emission detected by \citet{1994A&A...282..605J} and \citet{2008A&A...477..557M}, respectively.

Assuming that IC\,63 is located at its current projected distance of 1.2 pc, we obtained $\sim1050$ and $\sim0.4$ as the values for $\eta$ and $\nu$. Thus IC\,63 falls into the region II in figure 1 of \citet{1989ApJ...346..735B}. Clouds falling in this region are thought to be compressed by an ionization shock front \citep{1989ApJ...346..735B, 2005IAUS..235P.148P}. The shock advances into the cloud at a velocity of $v_{s}=\nu\,c_{i}$, where $c_{i}$ is the isothermal sound speed (11.4\,km\,s$^{-1}$, assuming an equilibrium temperature for the ionized gas of $T_{i}=10^{4}$ K). A steady flow of ionized gas streaming off the surface of the cloud creates an ionized boundary layer (IBL) which absorbs a fraction of the incident ionizing flux resulting in a reduction in the shock velocity. The photoevaporation parameter ($\psi$) introduced by  \citet{1989ApJ...346..735B} gives a  measure of the absorption efficiency of the ionized gas in the IBL. We estimated the value for $\psi$ using the expression given by \citet{1989ApJ...346..735B} as $\sim85$. Since $\psi\gg1$, the I-front is expected to be thin compared to the size of the cloud radius. The thickness of the I-front in IC\,63 is estimated as $\sim0.001$ pc. We estimated the I-front thickness by reading out the value of (1-$\phi_{q}$) from the figure 3 of  \citet{1989ApJ...346..735B} corresponding to the value of log($\psi$). Using equations 2.5 (gives a measure of $\nu$) and 6.2 (gives an estimate of the amount of incident ionizing flux get absorbed in the IBL that fumes out of the side of the cloud facing the ionizing source) \citep{1989ApJ...346..735B}, in a steady state condition, the shock velocity is found to be $\sim2$\,km\,s$^{-1}$. Thus the velocity attained by the cloud after its implosion is $v^{\prime}\simeq0.45\,v_{s}$ which is $\approx1$ \,km\,s$^{-1}$. 

The evolution of a cloud in the vicinity of massive stars is thought to occur in two main phases, namely a collapse and transient phase, and a cometary phase. The initial collapse phase is shown to last for $\sim10^{5}$\,yr followed by a relatively long typically $\sim10^{6}$\,yr cometary phase \citep{1994A&A...289..559L}. With the velocity derived above, the shock will take $\sim2\times10^{5}$\,yr to cross the length of IC\,63 which is $\sim0.2$ pc \citep{1991ApJS...77...59S} consistent with the time scale derived by \citet{1994A&A...289..559L}. The age of $\gamma$ Cas is estimated to be $\sim4$ Myr \citep{1997MNRAS.287..455B, 2005A&A...441..235Z} which suggests that IC\,63 is most likely in the cometary phase. The ionized gas that streams off the surface facing the ionizing star will escape with a mean velocity relative to the cloud. This process will impart an equal and opposite reaction on the cloud (similar to a jet from a rocket) and accelerate the cloud away from the ionizing star \citep{1955ApJ...121....6O}. \citet{1990ApJ...354..529B} estimated $c_{i}$ in terms of rocket velocity ($v_{R}$) assuming that it remains roughly constant as the cloud evaporates away from the ionizing star. We found $v_{R}\simeq0.8c_{i}$ which is $\sim9$\,km\,s$^{-1}$ corresponding to the values of log($\psi$) given by \citet{1990ApJ...354..529B}. Thus the net velocity that the IC\,63 could have gained because of the initial implosion and the acceleration due to the rocket effect is $\approx10$\,km\,s$^{-1}$. 

The radial velocity measurements of $\gamma$ Cas show a spread of data points in the approximate range of $-17$\,km\,s$^{-1}$ to 1\,km\,s$^{-1}$ \citep{2000A&A...364L..85H, 2012A&A...537A..59N}. If we take an average of the radial velocity measurements, then the value would be $-9$\,km\,s$^{-1}$. This is consistent with the systemic velocities derived by \citet{2012A&A...537A..59N} after the epoch of HJD 2452000. In IC\,63, the line emission peaks at a radial velocity of $\sim1$\,km\,s$^{-1}$ and in IC\,59, the peak emission is found at slightly higher at $\sim2$\,km\,s$^{-1}$. Thus the net radial velocity difference between the nebulae and the $\gamma$ Cas is found to $\sim10$\,km\,s$^{-1}$. Since acceleration due to the rocket effect depends only weakly on the parameter $\psi$, the velocity attained by the cloud will be of the same order that we calculated even in conditions where the actual initial separation between IC\,63 and $\gamma$ Cas were very small. 
%Given the current age of $\gamma$ Cas, the estimated net velocity will place IC\,63 about $\sim40$ pc away from the $\gamma$ Cas (assuming that IC\,63 was very close to $\gamma$ Cas to begin with). This will also place IC\,63 at an inclination angle of $\sim88\degree$ away from the plane of the sky. But the difficulty here is that this will also place the cloud out side of the Str{\"o}mgren radius ($\sim7$ pc, assuming an ambient density of 10 cm$^{-3}$ and a B0.5V spectral type) of $\gamma$ Cas. 

We note, however, that these values should be taken as an estimation only due to the uncertainties involved in the adopted quantities. For instance, compared to the values calculated by \citet{1996ApJ...460..914V}, \citet{2003ApJ...599.1333S} found the Lyman continuum photon emission rate to be less by a factor of $\sim1.5$ for a B0.5V spectral type star. In addition to this, a misclassification of half a spectral class may lead to an uncertainty in the predicted Lyman photon flux by a factor of two. Also, the actual separation between IC\,63 and $\gamma$ Cas will crucially depends on the estimated age of the $\gamma$ Cas which may also be quite uncertain. But even with such uncertainties involved in the analysis, dynamically, we expect the separation between IC\,63 and $\gamma$ Cas to be larger than the current projected distance of 1.2 pc. 
%If we assume that IC\,63 is located almost at the edge of the Str{\"o}mgren sphere, then the inclination angle become $\sim80\degree$ with respect to the plane of the sky. 

\subsubsection{The magnetic field strength}

Inclusion of a plane parallel or perpendicular magnetic field into the continuity equation across the I-front was shown to alter the propagation speed of I-front with respect to the material lying ahead of it when compared to non-magnetic scenarios \citep{1998A&A...331.1099R, 2000MNRAS.314..315W}. In 3D R-MHD simulations of the evolution of photoionized globule, \citet{2009MNRAS.398..157H} introduced magnetic fields of three different strengths and orientations. In the presence of a strong (186$\mu$G) perpendicular magnetic field, due to strong coupling, magnetic pressure and tension resist any movement of gas in the directions perpendicular to the original field lines. However, no such resistance is offered along the field lines resulting in a highly anisotropic evolution of the globule. In a strong parallel magnetic field orientation, the globule remains cylindrically symmetric throughout its evolution. The support provided by the parallel field lines resist any lateral compression which results in the formation of a broader, snubber globule head. In the case of an inclined magnetic field orientation, the cloud curls up to form a comma shape with the tail directed away but along the inclined field lines. For a weak and medium field strengths, initially perpendicular field lines are made to align along the direction of propagation of the ionizing photons \citep{2011MNRAS.412.2079M}.

We estimated the strength of magnetic field projected on the plane of the sky using a modified version of the classical method proposed by \citet{1953ApJ...118..113C}.  The modified equation, in terms of the molecular hydrogen density ($nH_{2}$), $\Delta$V the FWHM line width in km\,s$^{-1}$ and the dispersion in polarization angles ($\delta\theta$) in degree, obtained from \citet{2004ApJ...600..279C} is given as
\begin{equation}
B_{pos} = 9.3\sqrt{nH_{2}}\frac{\Delta V}{\delta\theta} \hspace{0.4cm}\mu\,G
\end{equation}
The original equation was derived assuming an equipartition between kinetic and perturbed magnetic energies. Together with the dispersion in polarization angles and in the  velocity along the line of sight, it was argued that the strength of the magnetic field projected on the sky could be estimated. To estimate the typical gas density towards the region sensitive to the optical polarization measurements, we first calculated the hydrogen column density using the relation, N(H$_{2}$) = 9.4$\times10^{20}\times$A$_{V}$ \citep{1978ApJ...224..132B}. We calculated an average A$_{V}$ of 1.9 magnitude using the A$_{V}$ values evaluated by \citet{2013ApJ...775...84A} of the six stars (\#5, 11, 52, 68, 75, 76) that are located within the cloud boundary. Assuming a line of sight thickness of the cloud to be similar to the width of IC\, 63 in the sky plane taken from \citet{1991ApJS...77...59S} for a distance of 215 pc ($\sim0.2$ pc), the hydrogen number density was found to be $3\times10^{3}$\,cm$^{-3}$. Using the CO (1-0) line width, $\Delta$V = 1.14 km\,$s^{-1}$ from \citet{1994A&A...282..605J} and the dispersion in polarization position angles after correcting for the position angle uncertainty (Lai et al. 2001; Franco et al. 2010), we obtained the plane of the sky magnetic field strength as $\sim90\mu$G. Magnetic field strengths in regions I, II and III towards IC\,63 (shown in Fig. \ref{Fig:combined_IC6359}) are estimated to be $\sim100\mu$G, $\sim65\mu$G, and $\sim90\mu$G, respectively. The uncertainty in the estimation of the number density is the dominant contributor to the uncertainty in the estimation of the magnetic field strength which could be as high as 40\%. The projected magnetic field estimated towards the outer regions of the cloud is not high enough to offer a full resistance to pressure due to the ionizing radiation. This explains the aligned field lines behind the condensation C\,1 and in the tail part of IC\,63.

%Because the cloud may be lying at an angle of  $\sim80\degree$ with respect to the plane of the sky, the actual orientation of the magnetic field with respect to the direction of the incoming ionizing radiation could be almost perpendicular as well.

Using the molecular density of the cloud, $(5\pm2)\times10^{4}$\,cm$^{-3}$, the line width (C$^{18}$O) of $(0.57\pm0.3)$ km\,$s^{-1}$ and temperature T$=50\pm10$\,K \citep{1994A&A...282..605J}, we derived thermal and turbulent pressure for molecular gas as $P_{th}$=$(3.5\pm1.6)\times10^{-10}$\,erg\,cm$^{-3}$ and $P_{turb}$=$(1.1\pm1.1)\times10^{-10}$\,erg\,cm$^{-3}$, respectively. The ionized gas pressure of the IBL in IC\,63 was estimated to be $2.7\times10^{-9}$\,erg\,cm$^{-3}$ by \citet{2004A&A...426..535M}. The dynamical pressure of the cloud therefore appears to be about twenty five times lower than the external pressure. In a scenario where the internal pressure is less than the pressure in the IBL, photoionization driven shock is expected to propagate into the molecular cloud causing it to implode and initiate star formation. However, there are no signs of star formation activity in IC\,63 \citep{2005AJ....129..954K} which implies that the cloud was not compressed enough during its dynamical evolution to become gravitationally unstable. The magnetic pressure, using the strength estimated towards the periphery of the cloud where the optical polarimetry is applicable, is found to be $P_{B}$=$(3.2\pm2.5)\times10^{-10}$\,erg\,cm$^{-3}$ which is also less by a factor of eight than the ionized gas pressure. For the magnetic pressure to provide support against the external pressure, the field strength would have to be greater than $\sim260\mu$G which is three times higher than the value estimated towards the outer parts of IC\,63. 

\subsection{IC\, 59}

\begin{figure}
%\resizebox{9.5cm}{8cm}{\includegraphics{IC59_only_fracp0.4_mod_W12.eps}}
%\resizebox{9.5cm}{8cm}{\includegraphics{\mypath/IC59_only_fracp0.4_3JUn16_W22.eps}}
\resizebox{9.5cm}{8cm}{\includegraphics{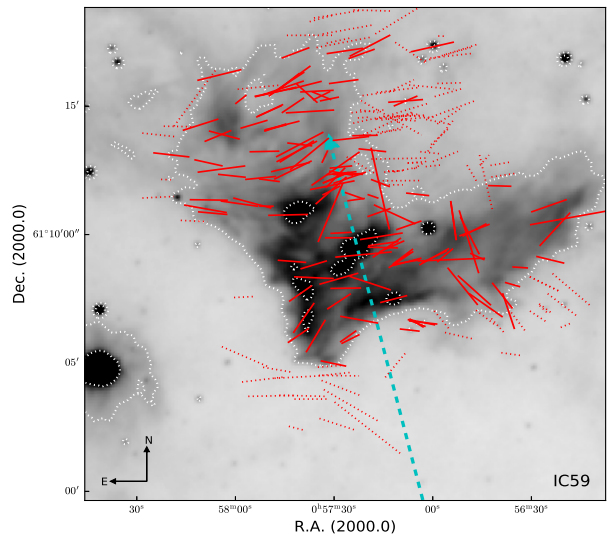}}
\caption{The polarization results of 236 stars observed towards IC\,59 region are shown on the 20$^{\prime}\times 20^{\prime}$ WISE 12$\mu$m image. The vectors drawn inside the outer most contour are shown using thick red lines. The ones outside this contour is shown using red dotted-lines. The broken arrow in cyan shows the direction of the ionizing photons from $\gamma~$Cas. The North and East directions are shown in the figure.}\label{Fig:IC59vectors}
\end{figure}

The plane of the sky component of the magnetic field in IC\,59 inferred from the polarization vectors are shown in Fig. \ref{Fig:IC59vectors}. Similar to IC\, 63, in IC\, 59 also, contours are drawn based on the 12$\mu$m intensity to define the cloud boundary. The contours are drawn at the levels of 0.29 and 1.89 Jy\,sr$^{-1}$. The morphology of this cloud does not belong to any of the three previously defined morphological types \citep[A, B or C; ][]{1991ApJS...77...59S} of BRCs. IC\, 59 displays a concave shaped surface towards the side facing the central star,  $\gamma$ Cas, which is located to the south of it. This concave surface is found to be located in between two cap features as identified and labeled in Fig. \ref{Fig:IC59vectors}. \citet{2010ApJ...717..658M} categorized this cloud as M-type because of its geometrical appearance. The symmetry lines running along the centres of the two caps are not radially directed towards the $\gamma$ Cas. Rather, it is the concave surface which is facing the illuminating star radially. The arrowed line in cyan in Fig.  \ref{Fig:IC59vectors} depicts the direction of the ionizing radiation ($\sim10\degree$ to the east from the north)  from $\gamma$ Cas. IC\, 59 lacks a prominent ionization rim which is why this cloud was not listed in the catalog of BRCs made by \citet{1991ApJS...77...59S}.  However, the H${\alpha}$ images of the region obtained by \citet{2010ApJ...717..658M} clearly show the presence of bright rims towards the two caps and the concave surface. The brightest rim is seen on the curved surface implying that stronger ionizing photons are received over there per unit area compared to surfaces of the two side caps. A CO condensation was found to be located behind the curved surface suggesting that the surface has undergone a compression due to the radiation impinging directly on it \citep{1998ApJS..115..241H}.

%-----------------------------------------------------------------
%-----------------------------------------------------------------

The peak emission in IC\,59 is found to be at a velocity redder than that of IC\,63 implying that IC\,59 is located at a farther distance than  IC\,63. The projected angular offset between the magnetic field and the direction of the ionizing radiation from $\gamma$ Cas is found to be $\sim80\degree$. The projected magnetic field geometry of IC\, 59 inferred from the polarization vectors lying outside the cloud boundary (region I of IC\,59 shown in Fig. \ref{Fig:combined_IC6359}) seems to be well ordered and oriented parallel to the ambient field direction or almost perpendicular to the direction of the ionizing radiation. But the field inside the cloud boundary (region II of IC\,59 shown in Fig. \ref{Fig:combined_IC6359}) seems to follow the shape of the concave surface towards the side facing the illuminating star. The histograms of $\theta_{P}$ corresponding to the two regions shown in red boxes towards IC\,59 and its distribution with degree of polarization are also shown in the right panels of Fig. \ref{Fig:combined_IC6359}. The dispersion of $\theta_{P}$ seen in these regions is relatively higher than that seen towards IC\,63 cloud. The magnetic field  seems to be oriented symmetrically about the direction of the ionizing radiation. Using SPH simulations, \citet{2010ApJ...717..658M} explained the formation of a M-type cloud morphology when exposed to the UV radiation from a nearby massive star. They followed the evolution of a uniform spherical molecular cloud (with an initial high ionization state) being exposed to high energetic radiation till a time of 1.7 Myr. At first, as the ionizing radiation compresses the star-facing side of the cloud, the spherical cloud evolves into a A-type  BRC. But because of the adopted very low initial density and the dependence of the shock velocity of the particles on their angular distance from the symmetry axis, the particles on the apex of the initially spherical cloud leads those lying away from the apex. Consequently, this results in the formation of a concave surface on the star-facing side of the molecular cloud and two cap structures on either side of it. A dense core starts to appear behind this surface as the cloud evolves further. The symmetry line of the cap structures becomes offset from the direction of the incident radiation. They showed that the current morphology of IC\, 59 is identical to the simulated structures at $t=1.11$ Myr. As the ionizing radiation penetrates more and more into the core, the temperature inside rises resulting in its expansion due to increase in the thermal pressure. Continuing the simulation further to $t=1.7$ Myr, they showed that the expanding cloud eventually becomes a large piece of a diffuse cloud. It is however interesting to note that the incoming ionizing radiation could not reorient the magnetic field along its direction implies that a stronger magnetic field strength prevails in IC\, 59. 

\section{Conclusions}\label{Conclusions}
We have derived the plane of the sky magnetic field geometry of two nebulae IC\,63 and IC\,59 which are associated with the H$~$II region, Sh 185 using R-band polarimetry of stars projected on the clouds. Based on the polarimetric results of stars from a region of $10\degree$ radius around the star $\gamma$ Cas, we estimated a distance of $215\pm40$ pc to Sh 185 and the nebulae. On a larger scale, we found that the magnetic fields of the two nebulae are anchored to the local interstellar magnetic field. Comparing the polarimetry to mid-infrared imaging from the WISE mission, we found two dense condensations  towards the head region of IC\,63 and the field lines are found to be aligned to both the head-tail morphology of the main condensation and to the direction of propagation of the ionizing radiation. The magnetic field geometry in IC\,63 is found to be oriented convex whereas in IC\,59 it is concave relative to the direction of the incoming ionizing radiation. The plane of the sky magnetic field strength in IC\,63 is found to be $90\mu$G. The magnetic field geometry of IC\,59 is found to follow the M shape structure of the cloud. We note that, based on the calculations by \citet{2010ApJ...717..658M}, the ionizing radiation is capable of modifying the magnetic fields through RDI induced evolution of the local molecular clouds. Comparing our observations to published theoretical models we find good general agreement, supporting the importance of magnetic fields in BRC evolution.

%=====================================================
%******************************************************************************************
%***********************ACKNOWLEDGEMENT****************************************************
%******************************************************************************************
\section{ACKNOWLEDGEMENT}
Authors thanks the referee for an encouraging report resulting significant improvement in the paper. We acknowledge the use of SIMBAD and NASA's \textit{SkyView} facility (http://skyview.gsfc.nasa.gov) located at NASA Goddard Space Flight Centre. C.W. L. was supported by Basic Science Research Program though the National Research Foundation of Korea (NRF) funded by the Ministry of Education, Science, and Technology (NRF-2016R1A2B4012593). A.S. thanks Piyush Bhardwaj for the help during the observations.

%\input{table4}

%\input{table5}

%======================================================================================
\bibliographystyle{mn2e}
\bibliography{IC6359ref}
%======================================================================================
%\clearpage
%======================================================================================

\label{lastpage}

\end{document}